\documentclass[longauth]{aa} 
\usepackage{graphicx}
\usepackage{natbib}
\usepackage[varg]{txfonts}
\usepackage{longtable}
\usepackage{txfonts}
\begin{document}
   \title{Optical Behavior of GRB~061121 around its X-Ray Shallow Decay
Phase}

   \subtitle{}
  \author{T.  Uehara\inst{1}  \and
          M.  Uemura\inst{2} \and
          A.  Arai\inst{3} \and
          R.  Yamazaki\inst{4} \and
          K.  S. Kawabata\inst{2} \and
          M.  Ohno\inst{5} \and
          Y.  Fukazawa\inst{1} \and
          T.  Ohsugi\inst{2}  \and
          M.  Yoshida\inst{2} \and
          S.  Sato\inst{6} \and
          M.  Kino\inst{6} 
          }

  \institute{
         Department of Physical Science, Hiroshima University,
         1-3-1 Kagamiyama, Higashi-Hiroshima 739-8526, Japan\\
         \email{uehara@hep01.hepl.hiroshima-u.ac.jp}
        \and
         Hiroshima Astrophysical Science Center, Hiroshima University, 1-3-1 Kagamiyama,
         Higashi-Hiroshima 739-8526, Japan
        \and
         Faculty of Science, Kyoto Sangyo University, Motoyama, 
         Kamigamo, Kita-Ku, Kyoto-City 603-8555, Japan
        \and
         Department of Physics and Mathematics, 
         Aoyama Gakuin University, 5-10-1 Fuchinobe, 
         Sagamihara 252-5258, Japan 
        \and
         Institute of Space and Astronautial Science, 
         Japan Aaerospace Exploration Agency, 3-1-1 Yoshinodai, 
         Chuo-ku Sagamihara, Kanagawa 252-5120, Japan
        \and 
         Department of Physics, Nagoya University, Furo-cho,
         Chikusa-ku, Nagoya 464-8602, Japan
}

   \date{Received ; accepted }

 
  \abstract
   {}
   { We report on a detailed study of the optical afterglow of GRB~061121 with 
     our original time-series photometric data.
 In conjunction with X-ray observations, we discuss the origin of its 
 optical and X-ray afterglows.
   }
   { 
 We observed the optical afterglow of {\it Swift} {\rm burst
 GRB~061121 with the Kanata 1.5-m telescope
 at Higashi-Hiroshima Observatory.  
 Our observation covers a period 
 just after an X-ray plateau phase.
 We also performed deep imaging with the Subaru telescope in 2010 
 in order to estimate the contamination of the host galaxy.
}
}
{
 In the light curve, we find that the optical afterglow also exhibited a
 break as in the X-ray afterglow.  
 However, our observation suggests a possible hump structure or a 
 flattening period before the optical break in the light curve.
 There is no sign of such a hump in the X-ray light curve.
}
   {
 This implies that the emitting region of optical was distinct  
 from that of X-rays.
 The hump in the optical light curve was possibly caused by the
 passage of the typical frequency of synchrotron emission from 
 another forward shock distinct from the early afterglow. 
 The observed 
 decay and spectral indices are inconsistent with the standard 
 synchrotron-shock model.  Hence, the observation requires a change in  
 microphysical parameters in the shock region or a prior activity 
 of the central engine.
 Alternatively, the emission during the shallow decay phase may be 
 a composition of two forward shock emissions, as indicated by 
 the hump structure in the light curve.
}


   \maketitle

\section{Introduction}
Gamma-ray bursts (GRBs) and their afterglows are widely believed to be
emission from relativistically expanding shells 
(e.g., \citealt{reviewZhang, Meszaros06}). 
GRBs, or prompt emissions are considered to
arise from internal shocks caused by collisions between the shells.  
After the collisions, the shell keeps expanding and generates an external shock
colliding with the interstellar medium.  
As a result, synchrotron emission from the shocked region 
is observed as afterglows.  
This synchrotron-shock
model successfully reproduces the observed temporal evolution of
spectral energy distributions (SEDs) of late afterglows.  
According to this model, the flux of the synchrotron emission from afterglows 
is described with a power-law form, that is, 
$f_{\nu}(t)\propto t^{-\alpha} \nu^{-\beta}$,
where $\alpha$ and $\beta$ are a decay index
and a spectral slope, respectively \citep{sar98grblc}.
A jet geometry was suggested by an achromatic break observed 
 in light curves of afterglows (e.g., Rhoads, 1997; 
 Sari et al., 1999). This has recently been called into 
 question because chromatic breaks were detected at the time 
 when the jet scenario predicts achromatic ones \citep{willingale07}.

Owing to quick identifications and notifications of GRBs by 
the {\it Swift} satellite, the number of observations 
of early afterglows has been
increasing in all wavelengths \citep{SWIFT}.  
X-ray light curves of early afterglows, in particular, 
turned out to have more complicate profiles than those previously expected 
from the standard synchrotron-shock model.  
Although a simple power-law decay was expected
in the standard model, the early X-ray light curves 
 actually consist of three stages with different decay indices; 
 the initial steep decay ($\alpha\sim 3-5$),
the shallow decay
($\alpha\sim 0.5-1.0$), and the normal decay phases ($\alpha\sim 1$)
(\citealt{nou06shallow, obr06shallow}).  
While the steep decay phase is likely 
a high-latitude emission of the prompt emission
(\citealt{kum00naked, yam06tail, lia06curvature, zha06swift}), 
the origin of the shallow decay phase is currently
unknown.  
Several models have been proposed for this phase, for
example, the late energy injection into the shocked region, the
time-dependent microphysics in the shock, or prior outflow emission
(\citealt{sar00refresh,
nou06shallow, zha01long, pan06shallow, 
tom06inhomo, iok06priact, dad06swift, yamazaki09}).

Each of those models predicts distinct behaviors of SED variations
in early afterglows.  Simultaneous multiwavelengths observations are
required to provide crucial clues on the nature of the early afterglow
phase.  \citet{pan06shallow} reported that the early 
break after the shallow decay phase is chromatic on the basis of
6 afterglows.  \citet{yos06prompt} additionally reported that
GRB~051109A also exhibited a clear chromatic break.
\citet{pan06shallow} proposed that the observed light curves require
the temporal evolution of microphysical parameters in the emitting 
region of early and late afterglows.
On the other hand, some afterglows apparently exhibited achromatic breaks after
the X-ray shallow decay phase.   
\citet{kruhler09} reported on optical---IR and X-ray light curves of GRB~080710, 
in which an achromatic break was observed.
\citet{blu06g050525a} reported another example of a possible achromatic break 
in GRB~050525A, while 
X-ray flares make difficult to accurately determine a break time
(also see, \citealt{klo05g050525a}).  In some of past cases, optical
observations were too sparse to determine break times and to catch
the detailed behavior on either side of the breaks.  
 We definitely need new observations in which break times can be determined 
 accurately both in X-ray and optical light curves.

GRB~061121 was detected by the {\it Swift} Burst Alert Telescope (BAT)
at 15:22:29~(UT) 21 November 2006 (\citealt{gcnr15}).  {\it Swift} also
reported the discovery of a bright optical afterglow 
with Ultraviolet/Optical Telescope (UVOT), which was
soon confirmed at 14.9~mag\footnote[1]{Mag show Vega magnitude in this paper.} 
by ground-based telescopes (\citealt{gcn5824}).
Its redshift was estimated to be $z=1.314$ by spectroscopic observations
of the optical afterglow (\citealt{gcn5826}).  This bright burst is a
typical GRB following a well-known empirical relationship between
$E_{\rm p}$ and $E_{\rm iso}$ (\citealt{gcn5848}).  The early X-ray light
curve has several breaks as other systems observed in the {\it Swift}
era (\citealt{nou06shallow}).  Prompt onsets of multiwavelengths
observations for GRB~061121 provided a unique opportunity to study the
temporal evolution of X-ray and optical afterglows (\citealt{gcn5824, 
gcn5827, gcn5828}).
 \citet{pag07GRB061121} reported on multiwavelength data during the prompt 
 and afterglow phase of this GRB.  According to them, both X-ray and optical 
 flux monotonically decayed, which can be described with an early exponential
 rise followed by a power-law decay phase.

Here we report on our optical and infrared observations using the Kanata
1.5-m telescope.  Our continuous time-series observations enabled us
to reveal the optical behavior near the X-ray shallow decay phase.  
We describe the details of our observations in section~2.  Combined
with other published data, we report the temporal
evolution of the optical and X-ray afterglows in section~3. 
In section~4, we discuss the nature of the variations in the light
curves using the synchrotron-shock model.  Finally, we summarize our
results in section~5.


\section{Observation and Data Analysis}
\subsection{Optical observations}

Our observation started at 16:37~(UT) 21 November 2006, 
$4.6\times 10^3\; {\rm s}$ after the GRB trigger time, and 
ended at 19:57~(UT).  The observation was
performed with TRISPEC attached to the Kanata 1.5-m telescope at
Higashi-Hiroshima Observatory of Hiroshima University.  TRISPEC is a
simultaneous imager and spectrograph with polarimetry covering both
optical and near-infrared wavelengths (\citealt{TRISPEC}).  
We used the imaging mode for the observation of GRB~061121 
and obtained 
77 sets of $R_{\rm c}$, $J$, and $K_{\rm s}$ band images.  
The exposure time of a $R_{\rm c}$-band image was 123~s.  
During the 123~s exposure, short exposures of a few seconds 
were taken for NIR arrays, and yielded net exposures of 
120 and 96~s for each $J$- and $K_{\rm s}$-band image, respectively.

The central wavelength of the TRISPEC's $R_{\rm c}$ system is $\sim 620\;{\rm nm}$, 
slightly shifted from the standard one ($=645\;{\rm nm}$).  
 The difference in magnitude between these systems is expected 
 to be 0.008~mag when a power law spectrum with a spectral 
 index of 1.0 is assumed.  In the following discussion, we neglect 
 this small difference.

\begin{figure*}
   \centering
   \includegraphics{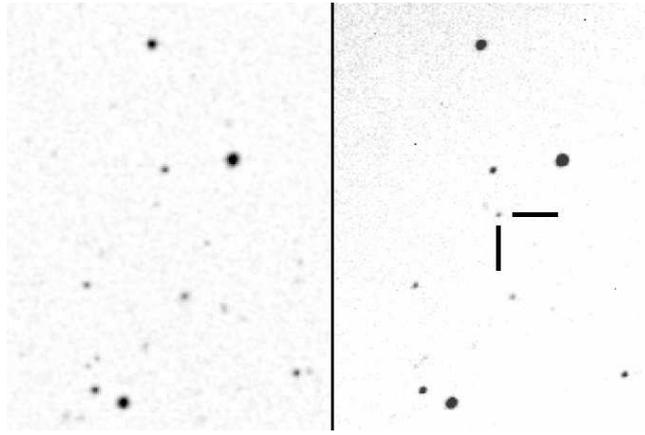}
   \caption{
    Optical images of the field of GRB~061121 in POSS2 (left
 panel) and observed with the Kanata 1.5-m telescope (right panel).  
 The field of view is $ 4\arcmin \times 3\arcmin$ and the top is north.  
 The afterglow is an
 object marked with the black bars in the right panel.
   }
              \label{fig:img}%
    \end{figure*}

We show an example of the obtained images in the right panel of
figure~\ref{fig:img}.  We also show the same field in the
Second Palomer Sky Survey (POSS2) in the left panel for comparison.  The
afterglow is the object marked with the black bars.  After making
dark-subtracted and flat-fielded images, we obtained magnitudes of the
afterglow and comparison stars using a Java-based PSF photometry
package.  
For a comparison star, 
we used USNO-B1.0 0768-0239968 (R.A.$=9^h48^m54^s.78$, 
Dec.$=-13\degr1\arcmin17\arcsec.9$; $R_{\rm c}=18.02$).
The comparison star was constant within
0.02~mag during our observation, 
checked by USNO-B1.0 0767-0229365 (R.A.$=09^h49^m05^s.080$, 
Dec.$=-13\degr13\arcmin22\arcsec.21$). 
Using neighbor USNO stars, we checked systematic errors of magnitudes
depending on comparison stars, and found that it is smaller than 0.2 mag.
 The comparison star is the same as that used in
 \cite{gcn5847}, \cite{gcn5851}, 
 and \cite{gcn5853} which present observations in a late stage of the afterglow.  
 In the following 
 section, we performed an analysis of our light curve in conjunction with 
 those late-time observations.

Additionally, table~\ref{tab:log} contains $R_{\rm c}$-magnitudes obtained by our
optical observations.  In this table, the magnitudes are averages in
equally spaced bins in the logarithmic scale of the time.  
While we obtained $J$ and $K_{\rm s}$ band images using
TRISPEC, IR afterglows were not significantly detected.  Typical 3-sigma
upper limits of each frame are 16.0 and 13.7~mag in $J$ and $K_{\rm s}$
bands, respectively.

%

   \begin{table}
      \caption[]{Results of our photometric observation.}
         \label{tab:log}
     $$ 
         \begin{array}{p{0.5\linewidth}l}
    \begin{tabular}{rcccc}
     \hline \hline
     Time (s)$^*$ & $R_{\rm c}^{**}$ mag. & $R_{\rm c}^\prime$ mag$^{\dag}$. & error & N$^{\ddag}$ \\
     \hline 
     4699 & 18.16 & 18.17 & 0.03 & 3\\
     5283 & 18.23 & 18.24 & 0.15 & 4\\
     6261 & 18.40 & 18.42 & 0.08 & 2\\
     6675 & 18.53 & 18.55 & 0.09 & 5\\
     7501 & 18.63 & 18.66 & 0.09 & 6\\
     8427 & 18.74 & 18.76 & 0.18 & 7\\
     9466 & 18.77 & 18.80 & 0.10 & 8\\
    10632 & 18.92 & 18.95 & 0.10 & 9\\
    11941 & 19.18 & 19.22 & 0.09 & 10\\
    13409 & 19.49 & 19.55 & 0.23 & 11\\
    15056 & 19.10 & 19.14 & 0.21 & 12\\
     \hline
    \end{tabular}
         \end{array}
     $$ 
\begin{list}{}{}
\item[*]  Time since the GRB trigger.
\item[**] Raw $R_{\rm c}$ magnitude.
\item[$\dag$]  $R_{\rm c}$ magnitude (host corrected).
\item[$\ddag$]  Number of images in each bin
\end{list}
   \end{table}

For deep photometry of the host galaxy
component, we obtained $R_{\rm c}$-band images with 
the 8.2-m Subaru Telescope and the Faint Object Camera 
and Spectrograph (FOCAS; \citealt{kashikawa02}) on
2010 May 7~(UT).
The total exposure time was 240~s. 
We can easily recognize the host galaxy as 
a point source at the GRB afterglow position in the 
obtained image.
 Using the same comparison star as mentioned above, we derived 
 the magnitude of the host galaxy to be $R_{\rm c}=22.99\pm 0.03$.

\subsection{Data analysis of the Swift data}

We analyzed public data of GRB~061121 observed with X-Ray Telescope
(XRT) and UVOT on {\it Swift}.  
We processed the XRT all orbits of data, adopting the standard
screening with the XRT pipeline FTOOL {\it xrtpipeline} (Version:~0.10.3).  
We extracted light curve and spectra with a
rectangular 40$\times$20-pixel region for the Windowed Timing (WT) mode,
and 40-pixel radius region for the Photon Counting (PC) mode from the
source position, respectively.  
The background was also extracted from 40$\times$20-pixel region 
for the WT mode, and 40-pixel radius region for the PC mode, 
far from the source, respectively.  While beginning of
GRB for WT mode data and PC mode data, we found that the count rate is
high enough to cause the pile-up effect, and we adopted the standard
pile-up correction as described by \citet{rom06pileup} and
\citet{vau06pileup}.

In the following section, the unit of time is set to be seconds from
the GRB trigger.  Optical and X-ray parameters are indicated by subscripts
of ``O'' and ``X'', respectively.  

\section{Results}

\subsection{Optical and X-ray light curves}

\begin{figure*}
\centering
\includegraphics[width=90mm, angle=270]{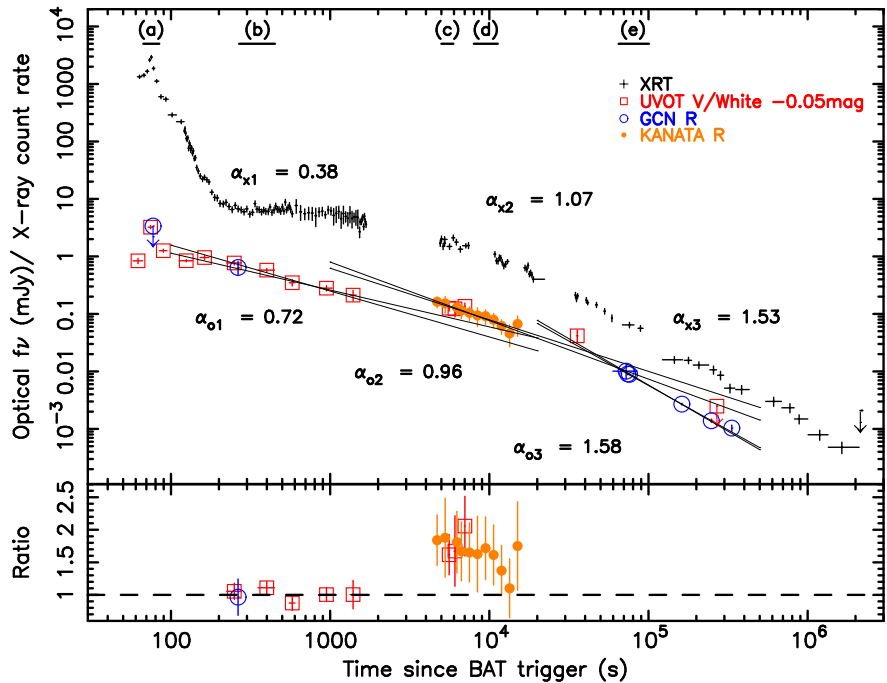}
\caption{
Optical and X-ray light curves of GRB~061121 afterglows.
The abscissa denote the time since the GRB trigger in seconds.  
In the top panel, the ordinate denote the flux density in mJy for optical data
and the count rate for X-ray data observed with XRT.  
 Our observations and X-ray observations by XRT are indicated by the filled
 circles and crosses, respectively. 
Open circles and squares
are optical observations with $R_{\rm c}$ and $V$/White light
(UVOT) bands reported to GCN or taken by UVOT, respectively.
The solid lines are 95~\% confidence regions
of the best fitted power-law models for optical light curves.
 The labels, (a),(b),(c),(d), and (e), represent the time intervals 
 for the SED analysis (see section~3.2).
 In the bottom panel, the ordinate denote the ratio of the observed flux density
 to the best-fitted model of the plateau phase.
$V$ band points are shifted by $-0.05$ to match the $R_{\rm c}$ band points.}

\label{fig:lc}
\end{figure*}

We show the X-ray and optical light curves of GRB~061121 in
figure~\ref{fig:lc}.  Our optical observations and X-ray observations by
XRT are indicated by the filled circles and crosses, respectively.  
The open circles and squares are optical observations with $R_{\rm c}$, 
$V$/``White light'' (UVOT) bands reported to GCN, respectively 
(\citealt{pag07GRB061121, gcn5824, gcn5827, 
gcn5830, gcn5833, gcn5837, gcn5840, gcn5844, gcn5847,
gcn5850, gcn5851, gcn5853, gcn5870}).  
About the optical flux, the contribution of the host galaxy is subtracted.
The flux density of the host galaxy is 2.14 $\mu Jy$ in $R_{\rm c}$, 
 which was estimated based on our 
 Subaru observation as described in section~2.
 The flux density was corrected for the Galactic extinction of 
 $E(B-V)=0.04$ (\citealt{sch98extinct}).
 The absolute magnitude of the host galaxy is $-21.86$~mag in $R_{\rm c}$.

According to \citet{pag07GRB061121}, the X-ray light curve of GRB~061121
is divided into 4 phases depending on their decay indices.  In this
paper, we follow their definition of the phases for the X-ray light
curves, that is, an initial flare, a plateau ($\alpha_{\rm X1}=0.38 \pm 0.06$), 
a shallow decay ($\alpha_{\rm X2}=1.07 \pm 0.05$), and a normal decay phases
($\alpha_{\rm X3}=1.53 \pm 0.03$). 
The errors of these parameters, as well as other parameters given 
 in this paper, represent 1-$\sigma$.

In the optical light curve, we can see a possible flare in a very early phase 
at $t=76\;{\rm s}$.  
This implies that the optical flux may be associated with 
the prompt emission in X-rays and $\gamma$-rays in this phase \citep{pag07GRB061121}.
 In the X-ray plateau phase, the optical light curve can be described with 
 a simple power-law decay.
Using $V$-band observations by UVOT from 240 to 2000~s, we calculated the power-law
decay index to be $\alpha_{\rm O1}=0.72 \pm 0.08$.

During the subsequent shallow decay phase, our observation revealed a
monotonic fading of the optical afterglow.  
The light curve can be described with a simple power-law having 
a decay index of $\alpha_{\rm O2}=0.96\pm 0.06$.  
 \citet{pag07GRB061121} reported a $V$-band decay index of $0.66\pm0.04$ 
 from the onset of the fading to a break at $\sim 2.5\times 10^4$~s.  
 This decay index was estimated based on the exponential--to--power-law model, 
 which assumes a monotonic fading during the fading stage of the afterglow.  
 The $\alpha_{\rm O2}$ estimated from our time-series photometry is, however, 
 significantly larger than that reported in \citet{pag07GRB061121}. 
We tried to fit a simple power-law model to the optical light curve
 from 240~s to 16~ks including our data.  
 For the fitting, our $V$-band data was shifted 
 to corresponding $R_{\rm c}$-band magnitude.  
 The $V-R_{\rm c}$ of the afterglow was estimated from two almost 
 simultaneous $V$- and $R_{\rm c}$-band observations, that is, 
 $t \sim280$ and $\sim6300\;{\rm s}$.  The average color of the 
 afterglow is calculated to be $V-R_{\rm c}=0.05$.  The best-fitted 
 parameters yield a chi-square/d.o.f of 76.3/15.
This value is too high to conclude that the afterglow decayed 
 with a simple power-law form from 240~s to 16~ks, and rather 
 suggests that there is a sub-structure around the termination
 of the X-ray plateau phase.

 The optical light curve, 
 then, exhibit another break around $3\times 10^4\,{\rm s}$, 
 which is followed by the normal decay phase described
with $\alpha_{\rm O3}=1.58\pm 0.03$.  
By fitting a broken power-law model, 
we calculated a break time of 
$4.6^{+4.5}_{-2.3}\times 10^4\,{\rm s}$. 
About the last break from the shallow decay to the normal decay phase, no
significant time lag is detected between the X-ray and optical breaks,
while the errors of break times are quite large;
$3.2^{+2.1}_{-0.6}\times 10^4\,{\rm s}$ and 
$4.6^{+4.5}_{-2.3}\times 10^4\,{\rm s}$ 
for the X-ray and optical break times, respectively.  
It is noteworthy that the optical decay index is almost same as the X-ray one
after this break.  According to the standard synchrotron shock model,
this strongly indicates the passage of the cooling frequency of the
synchrotron emission within the optical band at the break time.

\begin{figure*}
   \centering
   \includegraphics{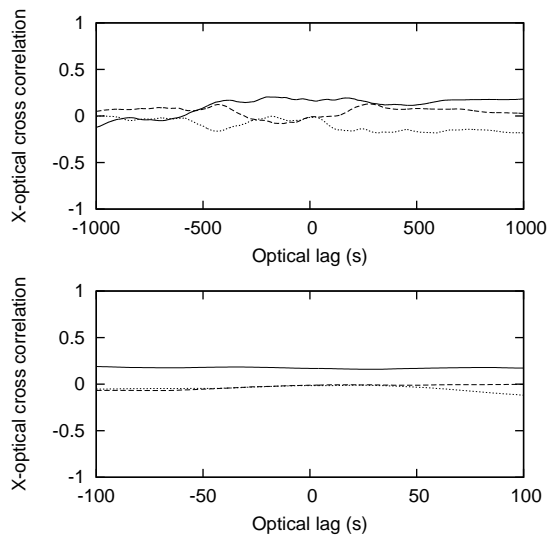}
   \caption{
X-ray/optical cross correlations for short-term variations.  The solid, dashed, and
 dotted lines were calculated from the data of 
 $4.8\times 10^3\;{\rm s}< t < 7.5\times 10^3\;{\rm s}$, 
 $1.0\times 10^4\;{\rm s}< t < 1.4\times 10^4\;{\rm s}$, and 
 $1.6\times 10^4\;{\rm s}< t < 1.9\times 10^4\;{\rm
 s}$, 
respectively. The
 upper and lower panels show correlation functions in long and short time scales,
 respectively.  No significant correlation can be seen.
   }
              \label{fig:corr}%
    \end{figure*}

We searched possible correlations between X-ray and optical short-term
variations.  We calculated cross-correlations using 3 segments 
in which simultaneous optical and X-ray data are available.  The optical 
data was divided into the following parts; 
i) $4.8\times 10^3\;{\rm s}< t < 7.5\times 10^3\;{\rm s}$, 
ii) $1.0\times 10^4\;{\rm s}< t < 1.4\times 10^4\;{\rm s}$, and 
iii) $1.6\times 10^4\;{\rm s}< t < 1.9\times 10^4\;{\rm s}$.
The resultant cross-correlations are shown in figure~\ref{fig:corr}.
The correlation functions are flat and show no prominent feature.  
We cannot detect any significant correlations between optical and X-ray
short-term variations.

\subsection{Spectral energy distribution} 

\begin{figure*}
   \centering
   \includegraphics{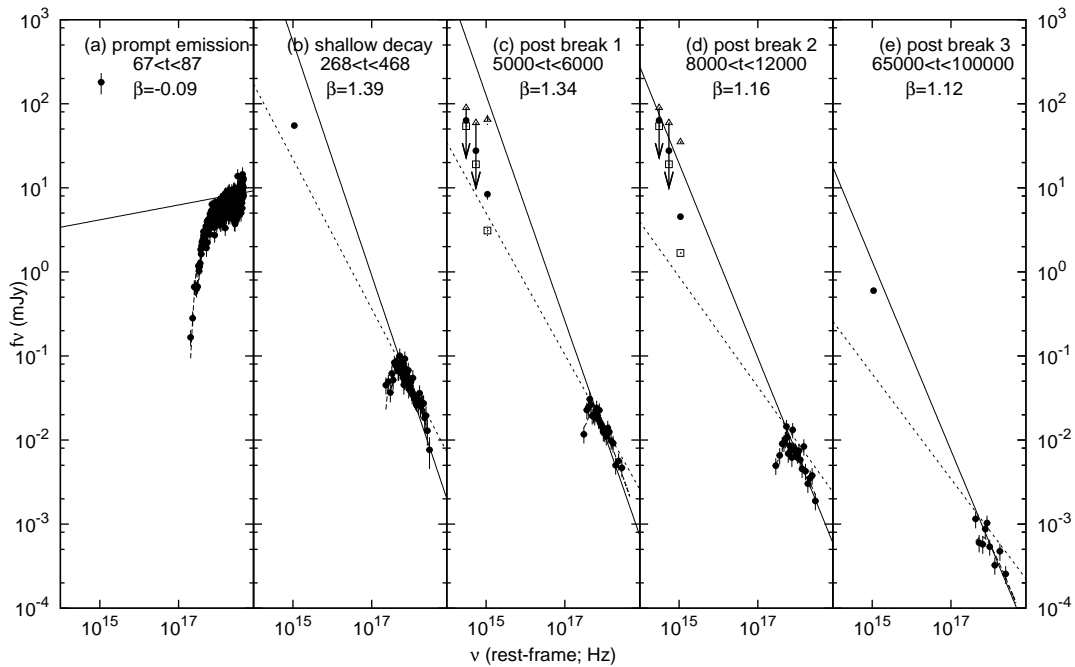}
   \caption{
Spectral energy distributions derived for the time intervals of 
(a) $6.7\times 10^1\;{\rm s}< t < 8.7\times 10^1\;{\rm s}$, 
(b) $4.3\times 10^2\;{\rm s}< t < 5.4\times 10^2\;{\rm s}$, 
(c) $5.0\times 10^3\;{\rm s}< t < 6.0\times 10^3\;{\rm s}$, 
(d) $8.0\times 10^3\;{\rm s}< t < 1.2\times 10^4\;{\rm s}$, 
 and (e) $6.5\times 10^4\;{\rm s}< t < 1.0\times 10^5\;{\rm s}$.  
 The time intervals of the SEDs for each panel are indicated in 
 figure~2.
 The abscissa and ordinate denote the rest-frame frequency in
 Hz and the flux density in mJy, respectively.  
 The filled circles are X-ray observations by XRT, 
 optical observations reported to GCN, and
 our optical---IR observations.  The IR observations just provide upper
 limits of the flux.  
 For the extragalactic extinction, the optical and IR data were 
 corrected with the ``Q2'' model (\citealt{mai01dust}).
 The open triangles and squares are data in which their corrections for the
 extinctions were performed with the Milky Way model and the ``Q1'' model, 
 respectively (\citealt{gui06extinct}).  
 We omit arrows indicating upper limits of the IR points of the open triangles
 and squares.  The solid lines represent the best fitted power-law model
 for the X-ray spectra.  The dotted lines are expected spectral slopes
 from the synchrotron-shock model.
   }
              \label{fig:sed}%
    \end{figure*}

Figure \ref{fig:sed} shows infrared--X-ray SEDs.  The figure contains 5
panels in which simultaneous optical and X-ray observations are
shown for the 5 phases.  
We fitted a power-law model with a single absorption component for the X-ray spectra.
All 5 spectra in figure~\ref{fig:sed} can be described
with absorption models 
with a hydrogen column density of 
$N_H=2.2 \pm $ 0.15 $\times 10^{21}\; {\rm cm}^{-2}$ 
in the observer's frame, which corresponds to 
$N_H=9.2\times 10^{21}\; {\rm cm}^{-2}$ in the rest-frame.  The solid
lines in the figure indicate the best fitted unabsorbed power-law
component of X-ray spectra.  
As can be seen from the figure, spectral
slopes $\beta$ ($f_\nu\propto \nu^{-\beta}$) were slightly larger in
the plateau phase (panel b) and a phase just after the
optical break (panel c) than those in later phases.
Table~\ref{tab:sed} contains results of our best fitted parameters in
each period.

In the figure, the optical flux was corrected for the galactic and
extragalactic extinctions.
 The correction for the extragalactic extinction was performed with 
 the relationship between the visual extinction $A_V$ and the hydrogen 
 column density $N_H$ for the ``Q2'' model in \citet{mai01dust} 
 ($N_{\rm H}/A_V=3.3\times 10^{21}\; {\rm cm}^{-2}$).
We estimated the $N_H$ from the best-fitted model of
X-ray spectra, and set to $N_H=9.2\times 10^{21}\; {\rm cm}^{-2}$ as
mentioned above.  
The conversion from $A_V$ to those in other bands was performed
following the equations in \citet{car89extinct}.

For the correction of the extragalactic extinction, we used the ``Q2''
model because it provides the most plausible optical---IR SEDs, as shown
below.  According to the synchrotron-shock model, the spectral slope at
the optical region should be $\beta_O=\beta_X-0.5$ in the case of 
$\nu_m < \nu_{\rm O} < \nu_c < \nu_{\rm X}$,
where $\nu_m$ and $\nu_c$ are typical and cooling frequencies of the
synchrotron emission from a forward shock (\citealt{sar98grblc}).  
 In figure~\ref{fig:sed}, we show the expected spectral slope of 
 $\beta_O=\beta_X-0.5$ with $\nu_c=10^{18}\;{\rm Hz}$, 
 indicated by the dotted lines.  
 The optical flux is required to be over this dotted lines 
 in order to satisfy the condition expected by the 
 synchrotron-shock model.
 In addition to the ``Q2'' model, the figure also contains 
optical--IR points corrected with the Milky Way
model ($N_{\rm H}/A_V=1.6\times 10^{21}\; {\rm cm}^{-2}$; open triangles)
and with the ``Q1'' model ($N_{\rm H}/A_V=6.7\times 10^{21}\; {\rm cm}^{-2}$; 
open squares)(\citealt{mai01dust}).  As can be seen in 
panel (c), a high $A_V$ provided by the Milky Way model yields an 
unnaturally sharp break between $R_{\rm c}$ and $J$-bands.  Corrected
by the ``Q1'' model, the optical flux is too low to be interpreted by
the synchrotron-shock model with $\beta_O=\beta_X-0.5$.  The models
for SMC ($N_{\rm H}/A_V=1.5\times 10^{22}\; {\rm cm}^{-2}$) and LMC
($N_{\rm H}/A_V=7.6\times 10^{21}\; {\rm cm}^{-2}$) also yield further
lower optical flux.  Thus, the ``Q2'' model provides the best 
correction among those models.  

Near the peak of the prompt emission, as can be seen in panel~(a)
of figure~\ref{fig:sed}, the optical flux is much above the power-law
component of X-rays.  This indicates that the emission mechanism or
source of the optical emission is distinct from those of the prompt
X-ray and $\gamma$-ray emission.

In panel~(c) of figure~\ref{fig:sed}, there is a difference between the
optical flux and the spectrum extrapolated from the X-ray data (the
solid line in the figure).  The SED, hence, requires a spectral break
between the optical and X-ray bands.  This is consistent with the
situation for the case of $\nu_m < \nu_{\rm O} < \nu_c < \nu_{\rm X}$.  
In the standard model, $\nu_c$ evolves with time, decreasing in case of 
 constant density medium, increasing in case of wind medium. 
 Since our findings privilege a constant density medium as discussed in 
 subsection~4.2, we expect at one point that $\nu_c$, decreasing with time, 
 will cross the optical band. At that time, the optical and X-ray decay 
 will become identical, and the SED will be compatible with a simple power 
 law. This is the case in panel (e), suggesting that $\nu_c$ crossed the 
 optical band around $t = 5 \times 10^4$~s.

\begin{table}
  \caption{Best fit parameters for X-ray spectra}\label{tab:sed}
  \begin{center}
    \begin{tabular}{ccc}
     \hline \hline
     Time interval (s)& $\beta$ & $\chi^2$/d.o.f\\
     \hline 
   $6.7\times 10^1< t < 8.7\times 10^1$ & $0.09\pm $ 0.01& 267/195\\
   $4.3\times 10^2< t < 5.4\times 10^2$ & $1.39\pm $ 0.04& 19/17  \\
   $5.0\times 10^3< t < 6.0\times 10^3$ & $1.34\pm $ 0.05& 14/12  \\
   $8.0\times 10^3< t < 1.2\times 10^4$ & $1.16\pm $ 0.04& 10/9   \\
   $6.5\times 10^4< t < 1.0\times 10^5$ & $1.12\pm $ 0.09&10/9   \\
     \hline
    \end{tabular}
  \end{center}
\end{table}

\section{Discussion} 

\subsection{A possible hump structure in the optical light curve}

Here, we discuss the substructure in the optical light curve around 5~ks.
The behavior of the optical and X-ray light curves is unclear in the
transition phase from the plateau to the shallow decay phases
($1.5\times 10^3\;{\rm s}\lesssim t \lesssim 4.6\times 10^3\;{\rm s}$).  
 In figure~\ref{fig:lc}, the solid lines denote 95~\% confidence 
 regions of the power-law decay model 
 for each phase of the optical afterglow.  
 As indicated by these lines, our 
 observation suggests that the plateau phase is 
 terminated by a flattening phase
 or a hump in the optical afterglow.

 We checked the significance of this hump with a correction of colors,
 because the best-fitted models were calculated with data taken with different 
 bands; the $V$- and $R_{\rm c}$-band observations 
 in the plateau and shallow decay 
 phases, respectively.  
 The lower panel of figure~\ref{fig:lc} shows the 
 ratio of the observations to the best-fitted model of the plateau phase.  
 We converted $V$-band observations to our $R_{\rm c}$-band ones, 
 by adding $V-R_{\rm c}=0.05$ to the $V$-magnitudes.
 The hump appears over a 2.2-sigma level even with the color
 correction.

 By contrast, there is no sign of such a  hump in the X-ray light curve 
 between those two phases.
 In addition, there is a large difference in the optical and X-ray decay 
 indices in the plateau phase, compared with those in the subsequent shallow 
 and normal decay phases.
 The X-ray and optical light curves apparently exhibit different
 behaviors during the X-ray plateau phase and the transition phase to the 
 shallow decay phase.

\subsection{Implication to the synchrotron-shock model}
In this subsection, 
we discuss the behavior of the afterglow of GRB~061121 based on
the synchrotron-shock model.  
In the case of
$\nu_m < \nu_{\rm O} < \nu_c < \nu_{\rm X}$, 
\citet{urata07} proposed a relation between the decay indices of the X-ray 
and optical bands described as $\alpha_{\rm X} - \alpha_{\rm O}= 1/4$.
In the plateau and shallow decay phases of GRB~061121, 
the $\alpha_{\rm X} - \alpha_{\rm O} $ is
$0.34 \pm$ 0.10 and $0.11 \pm$  0.08, respectively. 
The classical synchrotron-shock
model, hence, fails to reproduce the observed light curves of
GRB~061121 in the shallow decay phase.
It can only marginally reproduce the light curve in the plateau phase.

\citet{pan06shallow} generalized the formulae of the synchrotron-shock
model by including the variations of the energy ($E$) in the blast
wave, the energy ratio for electrons and the magnetic field ($\varepsilon_i$
and $\varepsilon_B$), and the ambient medium ($n$) in the following form; 
$E(>\Gamma)\propto\Gamma^{-e},\; \varepsilon_B\propto \Gamma^{-b},\;
\varepsilon_i\propto \Gamma^{-i},\;{\rm and}\; n(r)\propto r^{-s}$.  The
decay indices of optical and X-ray light curves are calculated as in
equations (9) and (10) in \citet{pan06shallow}.  Using those
formulae, we evaluate the presence of the energy injection ($e>0$) or
the time variations of microphysical parameters ($b\neq 0$ or 
$i\neq 0$) for GRB~061121.

We first assume $s=0$, namely the uniform distribution of the
interstellar medium.  In the following examination, we calculate $p$
from $\beta$ in each time-interval shown in table~\ref{tab:sed}.  For
the plateau phase, assuming $b=0$ and $i=0$, we find that the optical
and X-ray light curves yield inconsistent $e$, that is, $e=1.32\pm 0.24$
and $4.05\pm $ 0.37 calculated from $\alpha_{\rm O1}$ and 
$\alpha_{\rm X1}$, respectively.  This inconsistency can be reconciled
only when $s$ takes a narrow range of $s=1.22\pm 0.01$ and $e$ takes
an unnaturally large value ($e\sim 7$).  In the other case of $s>0$,
the inconsistency in $e$ becomes more extreme.  These results 
indicate that the observed light curves during the plateau phase cannot
be reproduced only with the energy injection scenario.  Temporal
variations of the microphysical parameters are, hence, required.
Assuming $e=0$ and $s=0$, we obtain $b=-2.10\pm 0.36$ 
and $i=2.04\pm$ 0.12.
The positive $i$ implies a low efficiency of the energy for accelerating
electrons in highly relativistic shocks.

During the shallow decay phase after the optical hump, the decay
indices changed to $\alpha_{\rm O2}=0.96$ and $\alpha_{\rm X2}=1.07$.
Even in this phase, the condition of $b=0$ and $i=0$ yields different $e$
values calculated from $\alpha_{\rm O2}$ (yielding $e=0.64\pm$ 0.20)
and $\alpha_{\rm X2}$ ($e=1.19\pm$ 0.25).  Assuming $e=0$, we obtain
$b=-0.50\pm$ 0.27 and $i=0.75\pm$  0.12 for this phase.  
 It is interesting to note that the absolute values of both $b$ and $i$ 
 decreases from the plateau to shallow decay phases.
 From the early to late stages, the condition of the blast wave
 may resemble the classical picture 
 in which no temporal variation in the microphysical parameters
 is required.
 Thus, the model proposed by \citet{pan06shallow} can explain 
 the light curves in both the plateau and shallow decay phases 
 by changing the microphysical parameters.

An alternative model was proposed by \citet{iok06priact}, which
consider prior activities before the main prompt emission.  
According to their model, the shallow decay of X-ray afterglows appears
because a blast wave obtains additional energy by colliding with
prior ejecta without significant decelerating of shells.  This model
is possibly preferable for GRB~061121, since it has a precursor 
$\sim 75\,{\rm s}$ before the peak of the main prompt emission
(\citealt{pag07GRB061121}).  The precursor may be a sign of the existence
of the prior activity.

\citet{iok06priact} define the the prior ejected mass as a power-law
form of $\gamma$, that is, $M(<\gamma)\propto \gamma^a$.  The decay
indices of X-ray and optical afterglows are given with 
$\alpha_{\rm X}=(a-3)/2 + (a-11)(p-2)/8$ and 
$\alpha_{\rm O}=(7a-25)/8 + (a-11)(p-2)/8$ in the case of 
$\nu_m < \nu_{\rm O} < \nu_c < \nu_{\rm X}$, respectively.  
For the plateau phase of GRB~061121, the X-ray and
optical decay indices provide a consistent $a$ within errors for
possible values of $p$ ($2.2\gtrsim p\gtrsim 2.8$).  In the case of
$p=2.78$, for example, 
we obtained $a=4.94\pm$ 0.48 from $\alpha_{\rm X1}$
and $a=5.05\pm$ 0.10 from $\alpha_{\rm O1}$.  
For the shallow decay phase, 
assuming $p=2.68$, for example, 
we calculated $a=5.99\pm$ 0.44 from $\alpha_{\rm X2}$ 
and $a=5.23\pm$ 0.10 from $\alpha_{\rm O2}$.
 Thus, the decay indices of the plateau and shallow decay phases can 
 be reproduced with the prior activity model proposed by \citet{iok06priact}, 
 while a temporal variation of $a$ would be needed.

 Both models in \citet{pan06shallow} and \citet{iok06priact} can
 explain the observed light curves of GRB~061121, only when time variations 
 in $b$, $i$, or $a$ are allowed.
The situation is further confusing when we consider
the presence of the optical hump between the plateau and shallow decay
phases.  The discontinuity around the hump indicates a variation of the
optical decay index, which means a further variation in $b$, $i$, or $a$ 
during the hump.

 The hump in the optical light curve between the plateau and shallow decay 
 phases is apparently not seen in the X-ray light curve.  
 This implies that the 
 dominant emitting regions are different in X-ray and optical 
 afterglows around the hump.  
 The hump structure reminds us of the two-component jet model; the hump may be 
 explained with a scenario that the emission from 
 a narrow jet may dominate in the plateau phase, 
 while that from a wide jet became dominant after the hump
 (\citealt{she03twojet, pen05twojet}).  In this case, the hump structure may 
 appear when $\nu_m$ of the synchrotron emission from the wide jet passes the 
 optical band.
 The classical synchrotron-shock model failed to reproduce the light curves 
 possibly because of the composition of the two components in the plateau and 
 shallow decay phases.

In the standard synchrotron-shock model, an increase of the density 
 of the shock region would produce a hump in the optical light curve 
 which is not seen in X-rays if $\nu_c$ lies between the optical 
 and X-ray bands (\citealt{pana00}). 
 Hence, the optical hump of GRB~061121 may also be reproduced if 
 the external shock passed through a high density region in the 
 interstellar medium at $t=2$--$6$~ks.

We finally note that there are several sources which exhibit optical light
curves analogous to GRB~061121, that is, GRB~021004 (\citealt{uem03g021004}), 
GRB~050525A (\citealt{klo05g050525a, blu06g050525a}), GRB~060117
(\citealt{jel06g060117}), GRB~060526 (\citealt{dai06g060526}), and GRB~061007
(\citealt{mun06g061007}). 
For all of them, an early decay phase was terminated by a flattening or
a hump at $10^{2-4}\;{\rm s}$, which was followed by a steeper decay
phase.  
An important point is that the decay indices before and after
the flattening phase are different each other.
This characteristic feature may commonly be observed
in a group of GRB afterglows.  If the emission during the early decay
phase has a different nature from that during the later decay phase, the
relationship of those two components possibly causes the diversity in
light curves of optical afterglows and the correlation between early
X-ray and optical light curves.  

\section{Summary}

 We performed time-series photometry of the optical afterglow of 
 GRB~061121 with the 1.5-m Kanata telescope, and reported on 
 a detailed study of the afterglow with published X-ray and optical 
 data.
 The decay index of the optical light curve was significantly different 
 between the plateau and shallow decay phases.
 The optical light curve possibly has a hump structure
 between the plateau and shallow decay phases, 
 while no sign of such a hump is
 seen in the X-ray light curve.  The different behavior in the
 optical and X-ray light curves indicates that they have
 distinct emitting sources.  
 The hump structure in the optical light curve may imply a passage of 
 the typical frequency of the synchrotron emission from another forward
 shock distinct from the early afterglow.
The optical decay index became same as
the X-ray one in the late phase after the final break at $\sim 4.6\times
10^4\,{\rm s}$.  In conjunction with the temporal evolution of SEDs,
we propose that this break is caused by the passage of the cooling
frequency at the optical band. 
 In both the plateau and shallow decay phases, 
 the observed decay and spectral indices 
 are inconsistent with the standard synchrotron-shock model.  
 They requires the variation of 
 microphysical parameters in the shock region or the prior activity 
 of the central engine.
It is also possible that they are due to the composition of two 
 forward shock components if the hump structure in the light curve was 
 caused by another forward shock.

\begin{acknowledgements}
The authors are grateful to Dr. M. Watanabe, for useful comments on
the paper.  This work was partly supported by a Grand-in-Aid from the
Ministry of Education, Culture, Sports, Science, and Technology of Japan
(17684004, 17340054, 19047004, 14079206, 18840032).  Part of this work
is supported by a Research Fellowship of Japan Society for the Promotion
of Science for Young Scientists.  
We are also grateful to K. Maeda, T. Hattori and M. Tanaka
for the opportunity of the Subaru observation for the
host component.
And, we are also grateful to Dr. Page, for useful UVOT data.
\end{acknowledgements}


\begin{thebibliography}{}

\bibitem[{Amati} et~al.(2006)]{gcn5848}
  {Amati}, L., {Frontera}, F., {Guidorzi}, C., \& {Montanari}, E.\ 2006, GRB
  Coordinates Network, 5848, 1

\bibitem[Bloom et~al.(2006)]{gcn5826}
  Bloom, J.~S., Perley, D.~A., \& Chen, H.~W.\ 2006, GRB Coordinates Network, 5826, 1

\bibitem[Blustin et~al.(2006)]{blu06g050525a}
  Blustin, A.~J., Band, D., Barthelmy, S., Boyd, P., Capalbi, M., Holland,
  S.~T., Marshall, F.~E., Mason, K.~O., {et~al.}\ 2006, \apj, 637, 901

\bibitem[Cardelli et~al.(1989)]{car89extinct}
  Cardelli, J.~A., Clayton, G.~C., \& Mathis, J.~S.\ 1989, \apj, 345, 245

\bibitem[Cenko(2006)]{gcn5844}
  Cenko, S.~B.\ 2006, GRB Coordinates Network, 5844, 1

\bibitem[{Dado} et~al.(2006)]{dad06swift}
  {Dado}, S., {Dar}, A., \& {De R{\'u}jula}, A.\ 2006, \apjl, 646, L21

\bibitem[Dai et~al.(2007)]{dai06g060526}
  Dai, X., Halpern, J., Morgan, N., Armstrong, E., Mirabal, N., Haislip, J.,
  Reichart, D., \& Stanek, K.\ 2007, ApJ 658, 509


\bibitem[Efimov et~al.(2006a)]{gcn5850}
  Efimov, Y., Rumyantsev, V., \& Pozanenko, A.\ 2006a, GRB Coordinates Network, 5850, 1

\bibitem[Efimov et~al.(2006b)]{gcn5870}
  Efimov, Y., Rumyantsev, V., \& Pozanenko, A.\ 2006b, GRB Coordinates Network, 5870, 1

\bibitem[Gehrels et~al.(2004)]{SWIFT}
  Gehrels, N., Chincarini, G., Giommi, P., Mason, K.~O., Nousek, J.~A., Wells,
  A.~A., White, N.~E., Barthelmy, S.~D., {et~al.}\ 2004, \apj, 611, 1005

\bibitem[Golenetskii et~al.(2006)]{gcn5837}
  Golenetskii, S., Aptekar, R., Mazets, E., Pal'shin, V., Frederiks, D., \&
  Cline, T.\ 2006, GRB Coordinates Network, 5837, 1

\bibitem[Guidorzi et~al.(2007)]{gui06extinct}
  Guidorzi, C., Gomboc, A., Kobayashi, S., Mundell, C.~G., Rol, E., Bode,
  M.~F., Carter, D., La~Parola, V., {et~al.}\ 2007, \aap, 463, 539

\bibitem[Halpern et~al.(2006b)]{gcn5840}
  Halpern, J.~P., Mirabal, N., \& Armstrong, E.\ 2006b, GRB Coordinates Network, 5840, 1

\bibitem[Halpern et~al.(2006a)]{gcn5847}
  Halpern, J.~P., Mirabal, N., \& Armstrong, E.\ 2006a, GRB Coordinates Network, 5847, 1


\bibitem[Halpern and Armstrong(2006a)]{gcn5851}
  Halpern, J.~P. \& Armstrong, E.\ 2006a, GRB Coordinates Network, 5851, 1

\bibitem[Halpern and Armstrong(2006b)]{gcn5853}
  Halpern, J.~P. \& Armstrong, E.\ 2006b, GRB Coordinates Network, 5853, 1




\bibitem[{Ioka} et~al.(2006)]{iok06priact}
  {Ioka}, K., {Toma}, K., {Yamazaki}, R., \& {Nakamura}, T.\ 2006, \aap, 458, 7

\bibitem[{Jel{\'{\i}}nek} et~al.(2006)]{jel06g060117}
  {Jel{\'{\i}}nek}, M., {Prouza}, M., {Kub{\'a}nek}, P., {Hudec}, R., {Nekola},
  M., {{\v R}{\'{\i}}dk{\'y}}, J., {Grygar}, J., {Boh{\'a}{\v c}ov{\'a}}, M.,
  {et~al.}\ 2006, \aap, 454, L119

\bibitem[Kashikawa et~al.(2002)]{kashikawa02}
Kashikawa, N. et~al.
\ 2002, \pasj, 819, 54

\bibitem[Klotz et~al.(2005)]{klo05g050525a}
  Klotz, A., Boer, M., Atteia, J.~L., Stratta, G., Behrend, R., Malacrino, F.,
  \& Damerdji, Y.\ 2005, \aap, 439, L35

\bibitem[{Kr{\"u}hler} et~al.(2009)]{kruhler09}
  {Kr{\"u}hler}, T. {Greiner}, J. \& {Afonso}, P. 2009, \aap, 508, 593


\bibitem[{Kumar}, {Panaitescu}(2000)]{kum00naked}
  {Kumar}, P. \& {Panaitescu}, A.\ 2000, \apjl, 541, L51


\bibitem[{Liang} et~al.(2006)]{lia06curvature}
  {Liang}, E.~W., {Zhang}, B., {O'Brien}, P.~T., {Willingale}, R., {Angelini},
  L., {Burrows}, D.~N., {Campana}, S., {Chincarini}, G., {et~al.}\ 2006, \apj,
  646, 351

\bibitem[Maiolino et~al.(2001)]{mai01dust}
  Maiolino, R., Marconi, A., \& Oliva, E.\ 2001, \aap, 365, 37

\bibitem[Marshall et~al.(2006)]{gcn5833}
  Marshall, F.~E., Holland, S.~T., \& Page, K.~L.\ 2006, GRB Coordinates Network, 5833, 1

\bibitem[Melandri et~al.(2006)]{gcn5827}
  Melandri, A., Guidorzi, C., Mundell, C.~G., Steele, I.~A., Smith, R.~J.,
  Monfardini, A., Carter, D., Kobayashi, S., {et~al.}\ 2006, GRB Coordinates Network, 5827, 1

\bibitem[M{\'e}sz{\'a}ros (2006)]{Meszaros06}
 {M{\'e}sz{\'a}ros}, P. 2006, Reports on Progress in Physics, 69, 2259


\bibitem[Mundell et~al.(2007)]{mun06g061007}
  Mundell, C.~G., Melandri, A., Guidorzi, C., Kobayashi, S., Steele, I.~A.,
  Malesani, D., Amati, L., D'Avanzo, P., {et~al.}\ 2007, \apj, 660, 489

\bibitem[Nousek et~al.(2006)]{nou06shallow}
  Nousek, J.~A., Kouveliotou, C., Grupe, D., Page, K.~L., Granot, J.,
  Ramirez-Ruiz, E., Patel, S.~K., Burrows, D.~N., {et~al.}\ 2006, ApJ, 642, 389

\bibitem[{O'Brien} et~al.(2006)]{obr06shallow}
  {O'Brien}, P.~T., {Willingale}, R., {Osborne}, J., {Goad}, M.~R., {Page},
  K.~L., {Vaughan}, S., {Rol}, E., {Beardmore}, A., {et~al.}\ 2006, \apj, 647,
  1213

\bibitem[Page et~al.(2006)]{gcnr15}
  Page, K.~L., Sakamoto, T., Marshall, F.~E., Barthelmy, S.~D., Burrows, D.~N.,
  Roming, P., \& Gehrels, N.\ 2006, GCN Report, 15.1

\bibitem[{Page} et~al.(2007)]{pag07GRB061121}
  {Page}, K.~L., {Willingale}, R., {Osborne}, J.~P., {Zhang}, B., {Godet}, O.,
  {Marshall}, F.~E., {Melandri}, A., {Norris}, J.~P., {et~al.}\ 2007, \apj,
  663, 1125

\bibitem[Panaitescu and Kumar(2000)]{pana00}
 {Panaitescu}, A. and {Kumar}, P. \ 2000, \apj, 543, 66

\bibitem[Panaitescu et~al.(2006)]{pan06shallow}
  Panaitescu, A., Meszaros, P., Burrows, D., Nousek, J., Gehrels, N., O'Brien,
  P., \& Willingale, R.\ 2006, \mnras, 369, 2059

\bibitem[Peng et~al.(2005)]{pen05twojet}
  Peng, F., Konigel, A., \& Granot, J.\ 2005, \apj, 626, 966

\bibitem[Piran(1999)]{pir99fireball}
  Piran, T.\ 1999, \physrep, 314, 575

\bibitem[Rhoads(1997)]{rho97break}
  Rhoads, J.~E.\ 1997, \apj, 487, L1

\bibitem[{Romano} et~al.(2006)]{rom06pileup}
  {Romano}, P., {Campana}, S., {Chincarini}, G., {Cummings}, J., {Cusumano},
  G., {Holland}, S.~T., {Mangano}, V., {Mineo}, T., {et~al.}\ 2006, \aap, 456,
  917

\bibitem[Rykoff et~al.(2006)]{ryk06g050801}
  Rykoff, E.~S., Mangano, V., Yost, S.~A., Sari, R., Aharonian, F., Akerlof,
  C.~W., Ashley, M. C.~B., Barthelmy, S.~D., {et~al.}\ 2006, \apj, 638, L5

\bibitem[Sari, Meszaros(2000)]{sar00refresh}
  Sari, R. \& Meszaros, P.\ 2000, \apj, 535, L33

\bibitem[Sari, Piran(1999)]{sar99reverse}
  Sari, R. \& Piran, T.\ 1999, \apj, 517, L109

\bibitem[Sari et~al.(1999)]{sar99break}
  Sari, R., Piran, T., \& Halpern, J.~P.\ 1999, \apj, 519, L17

\bibitem[Sari et~al.(1998)]{sar98grblc}
  Sari, R., Piran, T., \& Narayan, R.\ 1998, \apj, 497, L17

\bibitem[Schlegel et~al.(1998)]{sch98extinct}
  Schlegel, D.~J., Finkbeiner, D.~P., \& Davis, M.\ 1998, \apj, 500, 525

\bibitem[Sheth et~al.(2003)]{she03twojet}
  Sheth, K.~Frail, D.~A., White, S., Das, M., Bertoldi, F., Walter, F.,
  Kulkarni, S., \& Berger, E.\ 2003, \apj, 595, L33

\bibitem[Sonoda et~al.(2006)]{gcn5830}
  Sonoda, E., Maeno, S., Hara, T., Matsumura, T., Tanaka, K., Tanaka, H., \&
  Yamauchi, M.\ 2006, GRB Coordinates Network, 5830, 1

\bibitem[{Toma} et~al.(2006)]{tom06inhomo}
  {Toma}, K., {Ioka}, K., {Yamazaki}, R., \& {Nakamura}, T.\ 2006, \apjl, 640,
  L139


\bibitem[Uemura et~al.(2006)]{gcn5828}
  Uemura, M., Arai, A., \& Uehara, T.\ 2006, GRB Coordinates Network, 5828, 1

\bibitem[Uemura et~al.(2002)]{uem03g021004}
  Uemura, M., Kato, T., Ishioka, R., \& Yamaoka, H.\ 2002, \pasj, 55, L31

\bibitem[Urata et~al.(2007)]{urata07}
  Urata, Y., Yamazaki, R., Sakamoto T., \& Huang, K.\ 2007, \apjl, 668, L95




\bibitem[{Vaughan} et~al.(2006)]{vau06pileup}
  {Vaughan}, S., {Goad}, M.~R., {Beardmore}, A.~P., {O'Brien}, P.~T.,
  {Osborne}, J.~P., {Page}, K.~L., {Barthelmy}, S.~D., {Burrows}, D.~N.,
  {et~al.}\ 2006, \apj, 638, 920

\bibitem[Watanabe et~al.(2005)]{TRISPEC}
  Watanabe, M., Nakaya, H., Yamamuro, Y., Zenno, T., Ishii, M., Okada, M.,
  Yamazaki, A., Yamanaka, Y., {et~al.}\ 2005, \pasp, 117, 870

\bibitem[Willingale et~al.(2007)]{willingale07}
 Willingale, R., {et~al.}\ 2007, \apj, 662, 1093

\bibitem[{Yamazaki} et~al.(2006)]{yam06tail}
  {Yamazaki}, R., {Toma}, K., {Ioka}, K., \& {Nakamura}, T.\ 2006, \mnras, 369,
  311

\bibitem[{Yamazaki} (2009)]{yamazaki09}
  {Yamazaki}, R.\ 2009, \apjl, 690, L118


\bibitem[Yost et~al.(2006)]{gcn5824}
  Yost, S.~A., Schaefer, B.~E., \& Yuan, F.\ 2006, GRB Coordinates Network, 5284, 1

\bibitem[Yost et~al.(2007)]{yos06prompt}
  Yost, S.~A., Swan, H.~F., Rykoff, E.~S., Aharonian, F., Akerlof, C.~W.,
  Alday, A., Ashley, M. C.~B., Barthelmy, S., {et~al.}\ 2007, \apj, 657, 925

\bibitem[{Zhang} et~al.(2006)]{zha06swift}
  {Zhang}, B., {Fan}, Y.~Z., {Dyks}, J., {Kobayashi}, S., {M{\'e}sz{\'a}ros},
  P., {Burrows}, D.~N., {Nousek}, J.~A., \& {Gehrels}, N.\ 2006, \apj, 642, 354

\bibitem[{Zhang}, {M{\'e}sz{\'a}ros}(2001)]{zha01long}
  {Zhang}, B. \& {M{\'e}sz{\'a}ros}, P.\ 2001, \apjl, 552, L35

\bibitem[{{Zhang} \& {M{\'e}sz{\'a}ros}(2004)}]{reviewZhang}
{Zhang}, B. \& {M{\'e}sz{\'a}ros}, P. 2004, International Journal of Modern
  Physics A, 19, 2385

\end{thebibliography}
\end{document}